\documentclass[prl,reprint, amsmath,amssymb,aps, preprintnumbers,showpacs]{revtex4-1}
\usepackage[colorlinks=true,linkcolor=black, citecolor=black,urlcolor=black]{hyperref} 
\usepackage{graphicx,amsmath, amssymb, amsthm, amsfonts, mathtools}
\usepackage[caption=false]{subfig}
\usepackage{tikz}
\usetikzlibrary{arrows,positioning}
\usetikzlibrary{plotmarks}
\usetikzlibrary{patterns}

\newcommand{\ket}[1]{|#1\rangle}

\newcommand{\nonu}{\nonumber \\[2 mm]}
\newcommand{\be}{\begin{equation}}
\newcommand{\ee}{\end{equation}}
\newcommand{\bea}{\begin{eqnarray}}
\newcommand{\eea}{\end{eqnarray}}
\newcommand{\pat} {(\cdot 1 \cdot )}
\newcommand{\Kbar}{\overline{K}}

\date{v3, October 8, 2015}

\begin{document}
\preprint{OUTP-15-07P}
\title{Defects and degeneracies in supersymmetry protected phases}

\author{Thessa Fokkema$^{1}$}
\email{t.b.fokkema@uva.nl}
\author{Kareljan Schoutens$^{1,2}$}
\email{c.j.m.schoutens@uva.nl}
\affiliation{$^1$ Institute for Theoretical Physics, University of Amsterdam  Science Park 904, 1098 XH Amsterdam\\
$^2$ Rudolf Peierls Centre for Theoretical Physics, University of Oxford, 1 Keble Road, Oxford OX1 3NP}

\begin{abstract}
\noindent
We analyse a class of 1D lattice models, known as M$_k$ models, which are characterised by an order-$k$ clustering of spin-less fermions and by ${\cal N}=2$ lattice supersymmetry. Our main result is the identification of a class of (bulk or edge) defects, that are in one-to-one correspondence with so-called spin fields in a corresponding $\mathbb{Z}_k$ parafermion CFT. In the gapped regime, injecting such defects leads to ground state degeneracies that are protected by the supersymmetry. The defects, which are closely analogous to quasi-holes over the fermionic Read-Rezayi quantum Hall states, display characteristic fusion rules, which are of Ising type for $k=2$ and of Fibonacci type for $k=3$.
\end{abstract}

\maketitle

\vskip 3mm

{\bf Introduction.\ -}\ In the field of topological quantum computation (TQC) \cite{nayak:08}, a number of important lessons have been learned. The first is that non-Abelian statistics tend to be associated to a form of pairing or clustering in a quantum condensate. In 2D, $p$-wave pairing of spin-less fermions in a $p+ip$ superconductor or Moore-Read (MR) quantum Hall state gives rise to non-Abelian statistics of Ising type, through the mechanism of Majorana bound states at the cores of half flux quantum vortices or quasi-holes \cite{read:00, moore:91}. In the quantum Hall context, going beyond Ising anyons requires going beyond pairing, as in the Read-Rezayi (RR$_k$) \cite{read:98} and NASS$_k$ states \cite{ardonne:99}. The simplest examples beyond the MR state, the RR$_3$ and NASS$_2$ states, both give rise to Fibonacci anyons, which are universal for TQC (see, for example, \cite{ardonne:06}). More generally, the anyons carried by the RR$_k$ states are universal for $k=3$ and $k\geq5$ \cite{freedman:02, freedman:02_2}.

A second lesson learned is that a TQC-through-braiding protocol can be defined not just in 2D but also in a 1D setting \cite{alicea:11}. One starts from a T-shaped wire junction with non-Abelian defects at the wire ends and then runs a protocol of braiding, either in position space (by moving defects along the wires) or in parameter space. Again the prototypical example are Majorana bound states at the defect points. The underlying pairing is typically assumed to be extrinsic, meaning that it is induced (as in the Kitaev chain) through the proximity of a nearby superconductor.

We here consider the question if one can construct 1D lattice models with built-in, intrinsic, pairing or clustering properties and with defects binding Majorana or parafermion zero modes. We find that this goal is achieved by the lattice models M$_k$ introduced in \cite{fendley:03} and further analysed here. The definition of the M$_k$ models involves a hard-wired $k$-clustering constraint as well as ${\cal N}=2$ supersymmetry. The order-$k$ clustering leads to $\mathbb{Z}_k$ parafermion degrees of freedom in the CFT describing the M$_k$ models at criticality. In fact, the supercharge operator, which injects an extra particle into the system, contains the parafermion field $\psi_1$.  A similar structure is maintained if we move into a gapped phase. Our main result is the identification of a class of (bulk and edge) M$_k$ model defects that precisely correspond to the so-called spin fields $\sigma_i$ in the parafermion CFT. These defects are in many ways analogous to the quasi-holes over the RR$_k$ quantum Hall states: they have fractional particle number and display characteristic non-Abelian fusion rules. The underlying mechanism is that of supersymmetry protected order that is in essence of charge density wave (CDW) type. To turn this into supersymmetry protected topological order in the 1D sense will require a non-local reformulation via a Jordan-Wigner type transformation. 

\vskip 3mm

{\bf M$_k$ models.\ -}\ The M$_k$ lattice models \cite{fendley:03} describe spin-less fermions on a 1D lattice, subject to the `order-$k$ clustering' constraint that at the most $k$ particles can occupy consecutive sites. A supercharge $Q^+$ is defined as
\begin{equation}
\label{eqn:qplus}
Q^+ = \sum_{j=1}^L \sum_{a,b} \lambda_{[a,b],j} d^\dagger_{[a,b],j} ,
\end{equation}
where $d^\dagger_{[a,b],j}$ is a fermionic creation operator which creates a particle at lattice site $j$ in such a way that a string of $a$ particles is formed, with the newly created particle at position $b$. Choosing the $\lambda_{[a,b],j}$ such that $(Q^+)^2=0$, we define a ${\cal N}=2$ supersymmetric hamiltonian through
\begin{equation}
H= \{Q^+, Q^-\},
\end{equation}
with $Q^-=(Q^+)^\dagger$. This hamiltonian combines hopping terms with local potential and interaction terms. By construction, $[H,Q^+]=[H,Q^-]=0$. All states in the spectrum are doublets with $[f,f+1]$ particles, with the exception of the supersymmetric groundstates at $E=0$, which are annihilated by both $Q^+$ and $Q^-$. 

Possible choices for the $\lambda_{[a,b],j}$ have been studied in \cite{fendley:03, hagendorf:14}. Here we choose for the M$_2$ model
\begin{equation}
\lambda_{[1,1],j} = \sqrt{2} \, \lambda_j, \quad \lambda_{[2,1],j} = \lambda_{[2,2],j} = \lambda_j,
\end{equation}
with the $\lambda_j$ staggered as $\ldots 1 \lambda 1 \lambda 1 \ldots$. The factor $\sqrt{2}$ guarantees that the model is integrable \cite{hagendorf:14} and, if $\lambda=1$, critical \cite{fendley:03}.

For the general M$_k$ model we choose parameters describing a critical point perturbed by a specific, integrable, staggering \cite{hagendorf:15}. The staggering, with lattice periodicity $k+2$, connects the critical regime with one of `extreme staggering' $\lambda \ll 1$, where the $\lambda_{[a,b],j}$ follow a simple pattern. For $k=3$, to lowest order in $\lambda$,
\begin{equation}
\small{\begin{array}{ccccccccc}
\lambda_{[1,1],j}:& \ldots  & 1 & \sqrt{2}  & \sqrt{2} \lambda & \sqrt{2} & 1 & \ldots \\[1mm]
\lambda_{[2,1],j}:& \ldots & 1 & 1  & \lambda & \sqrt{2} & \lambda & \ldots \\[1mm]
\lambda_{[2,2],j}:& \ldots & \lambda & \sqrt{2} & \lambda & 1 & 1 & \ldots \\[1mm]
\lambda_{[3,1],j}:& \ldots & 1 & \lambda  & \lambda & 1 & \lambda/\sqrt{2}& \ldots \\[1mm]
\lambda_{[3,2],j}:& \ldots & \lambda/\sqrt{2} & 1 & \lambda^2/\sqrt{2}  & 1 & \lambda/\sqrt{2} & \ldots \\[1mm]
\lambda_{[3,3],j}:& \ldots & \lambda/\sqrt{2} & 1 & \lambda & \lambda & 1 & \ldots \\[1mm]
\end{array}}
\label{eq:stagk3}
\end{equation}
with the dots indicating repetition modulo 5. We denote this as $\ldots \star \star \lambda \star \star \ldots$, with the `$\lambda$' indicating the central position in the staggering pattern.

The Witten index for the M$_k$ model with periodic boundary conditions (PBC) and with $L=l(k+2)$ sites is $W_k=k+1$; indeed, for $\lambda>0$ the models have precisely this number of supersymmetric groundstates, all at $E=0$ and filling $\nu=k/(k+2)$ \cite{fendley:03, hagendorf:15b}. They are protected against perturbations that commute with supersymmetry and do not affect the $k$-clustering constraints. For open BC there are either zero or a single supersymmetric groundstate with $E=0$, the latter for $L \equiv 0,-1 \mod (k+2)$. We find, however, that in the presence of suitable boundary or bulk defects, the open systems have states with energies that are exponentially suppressed, $E\propto e^{-\alpha L_i}$, with $L_i$ characteristic distances among defects and boundaries. The exponential degeneracies are protected by supersymmetry.

\vskip 3mm

\begin{figure*}
\centering
   \subfloat[BC $open/open$]{
   \includegraphics[]{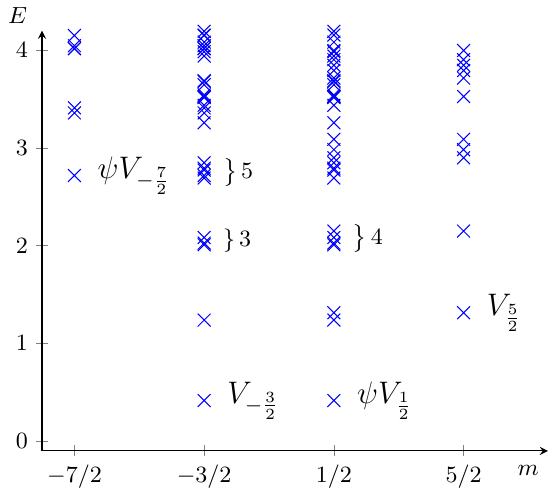}
       }
      \subfloat[BC $\sigma/open$]{
      \includegraphics[]{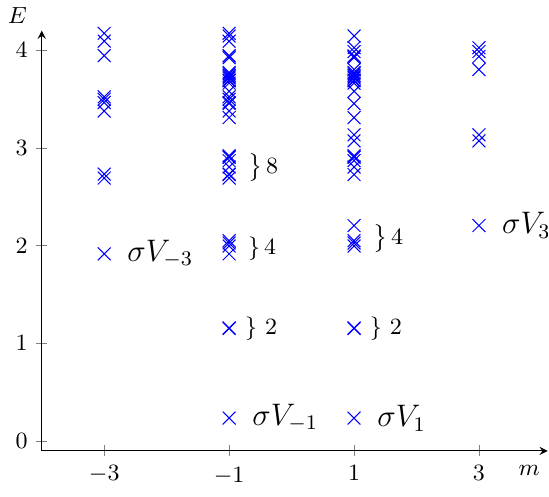}
      }
      \subfloat[BC $\sigma/ \sigma$]{
      \includegraphics[]{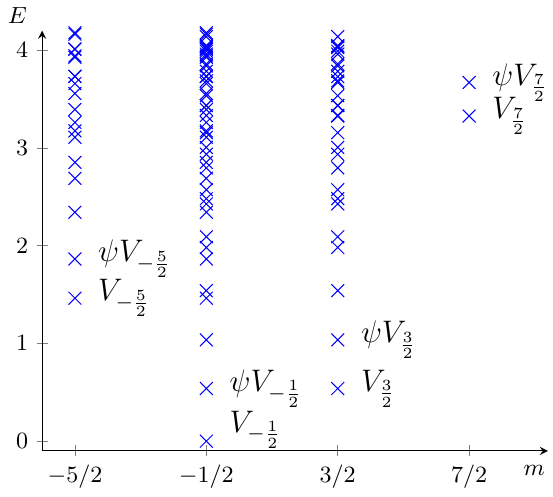}
       }
       \caption{Numerical M$_2$ spectra with  $L=25$, $f=11,12,13,14$ up to $E=4$. The labels specify the corresponding CFT modules.}
\label{fig:spectra}
\end{figure*}

{\bf M$_2$ model.\ -}\ We now zoom in on the M$_2$ model on an open chain. At criticality ($\lambda=1$), the finite size spectra can be matched with those of the 2nd minimal model of ${\cal N}=2$ superconformal field theory (CFT), of central charge $c={\frac{3}{2}}$. The match can be made with the help of numerical spectra (we analysed open chain spectra up to length $L=25$, fig.~\ref{fig:spectra}) and are similar to the results of \cite{huijse:11_2} for the M$_1$ model. The relevant CFT modules are $V_m$, $\psi V_m$ with $m \in \mathbb{Z} +{\frac{1}{2}}$ and $\sigma V_m$ with $m \in \mathbb{Z}$. Here the $V_m$ are charge $m$ vertex operators for a $c=1$ scalar field and the $\psi$, $\sigma$ arise from the $c={\frac{1}{2}}$ Ising CFT factor. States in $V_m$ have an even number of $\psi$-modes while those in $\psi V_m$ contain an odd number. The supercharges are the (Ramond sector) zero modes of the supercurrents ${\psi V_{\pm 2}}(z)$. 

Up to an overall $1/L$ scaling, the lattice model energies correspond to $E_{\rm CFT}=L_0-{\frac{1}{16}}$. The lowest energies are ${(4m^2-1)}/16$ for $V_m$, ${(4m^2+7)}/16$ for $\psi V_m$ and $m^2/4$ for $\sigma V_m$. The lowest-energy states in $V_{\pm \frac{1}{2}}$ and $\sigma V_0$ are supersymmetry singlets with $E_{\rm CFT}=0$, all other states have $E_{\rm CFT}>0$ and pair up into doublets. The critical $M_2$ spectrum with open BC  is easily described: with $m=2f-L-{\frac{1}{2}}$, one finds the CFT modules $V_m$ for $f$ even and $\psi V_m$ for $f$ odd (fig.~\ref{fig:spectra}). 

Reducing $\lambda$ below 1 sends the theory off criticality, with RG flow leading to the supersymmetric sine-Gordon (ssG) theory at coupling $\beta^2 = 8\pi$ \cite{hagendorf:14}.  The M$_2$ model off-critical finite size spectra can be analysed in terms of ssG  bulk $S$-matrices and boundary reflection matrices \cite{FS15}. The ssG theory holds important clues for the topological aspects of the M$_2$ model degeneracies \cite{bajnok:04, grosfeld:11, FS15}.

Rather than following the RG flow, we will here consider the limit $\lambda\ll 1$ (`extreme staggering'), where the M$_2$ eigenstates approach a simple factorized form. This is analogous to a special tuning in the Kitaev chain, which leads to perfectly decoupled Majorana edge states \cite{alicea:11}. This simple setting enables us to demonstrate how different BC result in exponential ground state degeneracy beyond this idealised limit. The $\lambda\ll 1$ limit is also similar to the thin-torus limit of the MR and RR$_k$ quantum Hall states \cite{bergholtz:06, seidel:06, ardonne:08}. Indeed, the systematics of the fusion channel degeneracies is highly analogous between the two settings. 

For $\lambda=0$ and PBC, the M$_2$ groundstates are 
\begin{eqnarray}
& |-\rangle =  \ldots 0\pat 0(\cdot 1 \ldots, \quad |+\rangle= \ldots 1\cdot)0 \pat 0 \ldots, \nonu
& |0\rangle = \ldots 1010101 \ldots,
\end{eqnarray}
where $\pat=110+011$ and $|-\rangle$ and $|+\rangle$ are related by a shift over two lattice sites. For open BC, there is at the most a single $E=0$ groundstate for given particle number $f$. For $L=4l-1$, staggering $1\lambda \ldots \lambda 1$, $f=2l$,
\begin{equation}
| + \rangle_{\rm o,o} = [\pat 0 \pat \ldots \pat] \ ,
\label{eq:oostate}
\end{equation}
where `o,o' refers to open/open BC. For $\lambda>0$ this state remains at $E=0$, where it is protected by the Witten index, $W=1$, and it is separated from all other states by a gap that remains finite as long as $\lambda<1$. 

\vskip 3mm

{\bf Boundary defects.\ -}\ To steer into a case with exponentially degenerate groundstates at given particle number $f$, we need to enforce a defect at both boundaries that allows all three $\lambda=0$ PBC groundstates to connect to the edge at zero energy cost. For this we impose the constraint that the two sites adjacent to a boundary cannot both be occupied by a particle \footnote{Similar BC were introduced independently in \cite{hagendorf:15b}}. With this BC (which we call of `$\sigma$-type' and denote by a bracket $   \ldots ]_\sigma$), all three $\lambda=0$ PBC vacua can connect to the boundary at zero energy cost. This gives, for $L=4l+1$, $\lambda1 \ldots 1\lambda$,
\begin{eqnarray}
& |-\rangle_{\sigma,\sigma} = {}_\sigma[ 0\pat \ldots 0]_\sigma, 
\quad |+\rangle_{\sigma,\sigma} = {}_\sigma[ 100 \pat \ldots 001]_\sigma, 
\nonu 
& |0\rangle_{\sigma,\sigma} = {}_\sigma[ 1010 \ldots 0101]_\sigma .
\label{eq:topostates}
\end{eqnarray}
For $\lambda>0$ the state $|- \rangle_{\sigma,\sigma}$ at $f=2l$ remains at $E=0$ while $|+\rangle_{\sigma,\sigma}$ at $f=2l$ and $|0\rangle_{\sigma,\sigma}$ at $f=2l+1$ pair into a doublet of energy $\delta E(\lambda)>0$. The energy difference between $| \pm \rangle_{\sigma,\sigma}$ originates from the boundary, which implies that it involves a power of $\lambda$ that scales with the length of the system. We checked numerically that
\begin{equation}
\delta E(\lambda) \propto \lambda^{(L-1)/2} , 
\end{equation}
which gives exponential splitting $\delta E\propto e^{-\alpha L}$ as long as $\lambda<1$. We checked that this behaviour is robust against perturbations obtained by deforming some of the parameters $\lambda_{[a,b],j}$ in eq.~(\ref{eqn:qplus}), provided we do not break the $\mathcal{N}=2$ supersymmetry. We thus identify supersymmetry as the agent protecting the exponential degeneracy of our `qubit' $| \pm \rangle_{\sigma,\sigma}$. We remark that many `natural' perturbations of the M$_2$ model do break supersymmetry. An example is the local fermion density operator $\rho_i$ at site $i=4p+1$. At $\lambda=0$ the expected value is  $\langle \rho_i \rangle =0$ in the state 
$ |-\rangle_{\sigma,\sigma}$ while $\langle \rho_i \rangle =1$ in the state $|+\rangle_{\sigma,\sigma}$.

\begin{figure}
\begin{center}
\includegraphics[width=0.49\textwidth]{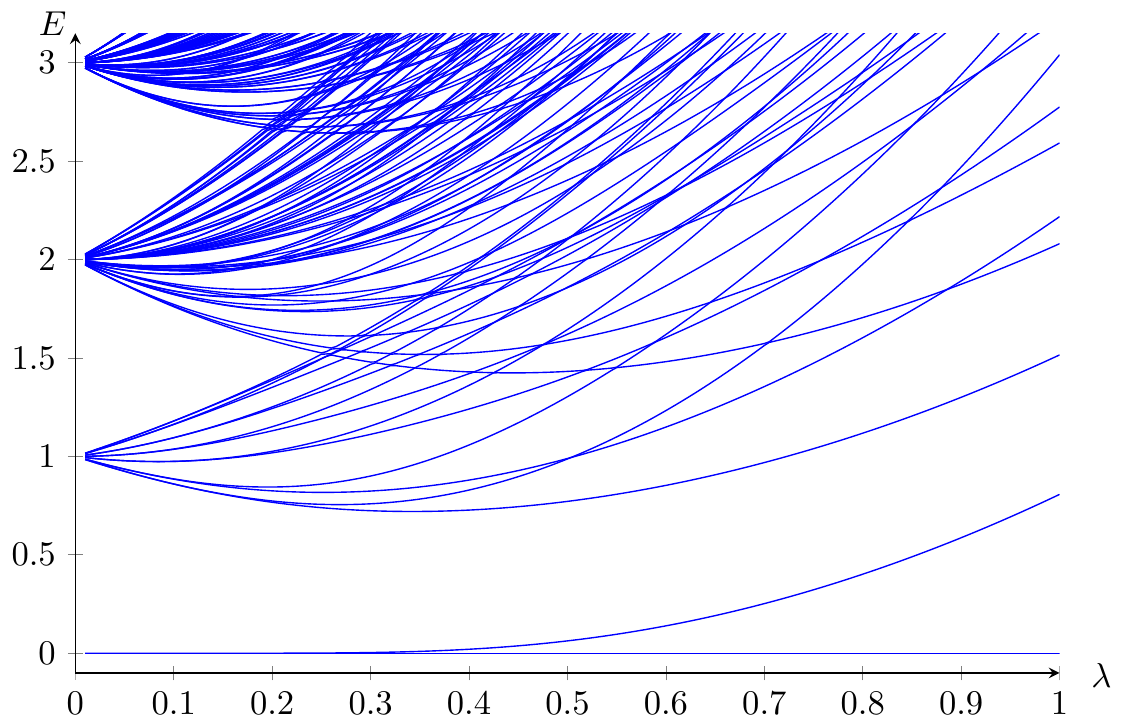}
\end{center}
\vspace{-5mm}
\caption{\small{Flow between `extreme staggering' (left, $\lambda=0$) and critical (right, $\lambda=1$) limits of the M$_2$ model with $L=17$ sites, $f=8$ particles, staggering $\lambda1 \ldots 1\lambda$ and $\sigma$/$\sigma$ BC. The lower two states constitute the `qubit' eq.~(\ref{eq:qubit}).}}
\label{fig:flow}
\end{figure}

Following the `qubit' states all the way to the CFT point, $\lambda=1$, we find (fig.~\ref{fig:flow})
\begin{equation} 
| - \rangle_{\sigma,\sigma}\leftrightarrow |V_{-\frac{1}{2}}\rangle, \quad
| + \rangle_{\sigma,\sigma}\leftrightarrow  |\psi V_{-\frac{1}{2}}\rangle,
\label{eq:qubit}
\end{equation}
where $|\cdot \rangle$ denotes the lowest weight state of the corresponding CFT module. At the CFT point the boundary Majorana modes have delocalised and the energy splitting is of order $1/L$.

The degeneracy of the `qubit' $| \pm \rangle_{\sigma,\sigma}$ can be traced to the fusion channel degeneracy $\sigma \sigma = 1 + \psi$ of the Ising spin-field $\sigma$ in the underlying CFT. Through the qH-CFT connection \cite{moore:91}, this choice of fusion channel carries over to the fusion product of two quasi-holes over the MR state. We consider the MR state in spherical geometry, which we view as an open `tube' capped by specific boundary conditions at the two poles. For $N=2l$ particles, the MR groundstate has the following thin-torus form
\begin{equation}
{\rm MR}, N=2l: \ \  [11001100 \ldots 110011] \ .
\end{equation}
The analogous groundstate of the M$_2$ model is precisely the state  $|+\rangle_{\rm o,o} $ in eq.~(\ref{eq:oostate}). The simplest case with two-fold fusion channel degeneracy is that of the MR states with $n=4$ quasi-holes. The general counting formula for $n$ quasi-holes reads \cite{read:96}
\begin{equation}
\sum_{F\equiv N \bmod 2} \left( \begin{array}{c} {\frac{N-F}{2}} + n \\ n \end{array} \right) \left( \begin{array}{c} n/2 \\ F \end{array} \right) .
\label{eq:MRqhcounting}
\end{equation}
Here the first binomial counts orbital degeneracies of the $n$ quasi-holes, while the second, together with the sum over $F$,  pertains to the fusion channel degeneracy. We fix the orbital degeneracy by selecting the states with two quasi-holes at both the north and the south poles,
\begin{equation}
F=0:\ [01100 \ldots 110] , \quad F=2:\ [10011 \ldots 001] .
\end{equation}
These `MR qubit' states correspond to the `M$_2$ qubit' states $| \pm \rangle_{\sigma,\sigma}$ of eq.~(\ref{eq:topostates}). 
To understand this we have to compare the open M$_2$ chains with the MR states not on the sphere but on the cylinder.  On a cylinder with vacua `1100' at far left and right, we can extend the $F=0$ state as
\begin{equation}
\ldots 1100 {}_{\sigma\sigma}[01100 \ldots 110 ]_{\sigma \sigma} 00 11 \ldots,
\end{equation}
where $\sigma \sigma$ denote the two quasi-holes at the boundaries. We can move one of the quasi-holes out from each of the boundaries to get
\begin{equation}
\ldots 1100 {}_{\sigma} 1010\ldots 10 {}_\sigma [01100\ldots110]_{\sigma} 0101\ldots 01 {}_\sigma 0011\ldots.
\end{equation}
This corresponds to the situation that we have in the M$_2$ model, where $\sigma$-type BC arise from the presence of a single $\sigma$ quantum at a boundary. This interpretation is confirmed by the CFT content of the open chain finite size spectra at $\lambda=1$ (fig.~\ref{fig:spectra}). 
For open/$\sigma$ BC, we find all sectors $\sigma V_m$ with $m$ shifted to $m=2f-L$. For $\sigma$/$\sigma$ BC, with non-Abelian defects at both ends, we find, at $m=2f-L+{\frac{1}{2}}$, both the $V_m$ and $\psi V_m$ modules, in accordance with the fusion rule $\sigma \sigma=1+\psi$. The states in eq.~(\ref{eq:qubit}) are a particular example.

\vskip 3mm

{\bf Bulk defects and quantum register.\ -}\  At extreme staggering, kinks connecting any two of the $\lambda=0$ vacua come at finite energy cost. For example, kinks/anti-kinks connecting $|0\rangle$ and $|+\rangle$, written as 
\begin{equation}
\ldots 10101\overbracket{\,0\,}^K 0\pat \ldots, \quad \ldots 10101\overbracket{\,1\,}^{\Kbar} 0\pat \ldots,
\label{eq:KKbar}
\end{equation}
have $E=1$; the same holds true for all kink types $(a,b)=(0,\pm),(\pm,0)$.  Note that kinks/anti-kinks are connected by ${\cal N}=2$ supersymmetry,
\begin{equation}
Q^+ : K_{a,b} \rightarrow \Kbar_{a,b}.
\end{equation} 
Multi-kink/anti-kink states can be counted through formulas similar to those for the MR state, see eq.~(\ref{eq:MRqhcounting}), for all choices of open chain BC. Followed through to the CFT limit, these counting formulas provide novel expressions for characters of the CFT \cite{FS15}.

We now define $\sigma$-type bulk defects. These will allow some of the bulk kink states to exist at zero energy (for $\lambda=0$).  In the example of eq.~(\ref{eq:KKbar}) this can be done by excluding the configuration `11' at the kink location: this eliminates the anti-kink and turns the kink into an $E=0$ state! To treat the $\pm$ vacua on equal footing, we repeat the same constraint two steps to the right, and define
\begin{equation}
\sigma :  \ldots \underbracket{\lambda\ 1}_{{\rm no}\ 11}\ \underbracket{\lambda\ 1}_{{\rm no}\ 11}\ \ldots, \quad
\sigma^\prime:  \ldots \underbracket{1\ \lambda}_{{\rm no}\ 11}\ \underbracket{1\ \lambda}_{{\rm no}\ 11}\ \ldots .
\end{equation}
This definition holds for general $\lambda$ and represents a constraint in the Hilbert space of states. At $\lambda=0$, a defect $\sigma$ can connect a vacuum $\ket{0}$ coming in from the left to $\ket{+}$, $\ket{0}$ or $\ket{-}$ extending to the right, and opposite for $\sigma^\prime$, all at zero energy.  

We can now conceive a `quantum register' by taking an open chain, length $L=4l+1$, staggering type $\lambda 1  \ldots 1 \lambda$, $\sigma$-type BC at both ends, and injecting a sequence of $2n$ well-separated defects in the order $\sigma^\prime \sigma \ldots \sigma^\prime \sigma$. This leads to $3^{n+1}$ degenerate groundstates at $\lambda=0$, with $2^{n+1}$ of them having the minimal particle number $f=2l-n$. These $2^{n+1}$ states form an Ising anyon `quantum register'. At $\lambda>0$, a unique $E=0$ groundstate at $f=2l-n$ remains, while all other states pair up into doublets at energies of order $e^{-\alpha L_i}$.  

\vskip 3mm

{\bf M$_3$ model.\ -}\ Turning to $k=3$ we identify the following four PBC groundstates at extreme staggering 
\begin{eqnarray}
& | 1 \rangle =  \ldots 1 \underset{\wedge}{1} 100\ldots, \
& | \tfrac{1}{2} \rangle =  \ldots (\cdot \underset{\wedge}{1} \cdots) \ldots,
\nonu
& | 0 \rangle = \ldots 0 \underset{\wedge}{1} 0 11 \ldots, \
& | -\tfrac{1}{2} \rangle = \ldots  \underset{\wedge}{0} (\cdot 1 1\cdot) \ldots, 
\end{eqnarray}
 with ${}_{\wedge}$ indicating the position of `$\lambda$' in the staggering pattern eq.~(\ref{eq:stagk3}), $(\cdot 1\cdots)=01101-01110+11001-11010$ and $(\cdot 1 1 \cdot)=1110-0111$. The energies of the kinks $K_{a,b}$ are $m_{a,b}= 2|a-b|$. This is in agreement with the kink masses in the ${\cal N}=2$ supersymmetric massive integrable QFT with Chebyshev superpotential $W(X)=\frac{1}{5} X^5- \beta^2 X^3 +  \beta X$ \cite{dijkgraaf:91, cecotti:91, gepner:90, fendley:91}, which we expect to result from the RG flow set by the staggering perturbation.

On open chains, we define $\sigma_1$ type BC by excluding `111' near a given edge, and $\sigma_2$ type by excluding `11'. At the level of the open-chain CFT spectra, a $\sigma_i$ type BC precisely corresponds to the $\mathbb{Z}_3$ parafermion spin field $\sigma_i$: upon changing BC, the various CFT sectors shift according to the fusion products with these two spin fields. The $k=3$ CFT contains, besides a free boson, $\mathbb{Z}_3$ parafermions $\psi_{1,2}$ and parafermion spin fields $\sigma_i$, $\varepsilon$. The supercharge $Q^+$ is the zero-mode of the super current $\psi_1 V_{\frac{5}{2}}(z)$. For open/open BC, the M$_3$ model realises the sectors, with $m=\frac{5}{2}f-\frac{3}{2}L-\frac{3}{4}$, 
\begin{equation}
\{ V_m,  \psi_1 V_m, \psi_2 V_m \}
\end{equation}
for $k=0,1,2$ with $k \equiv f \mod 3$.  For open/$\sigma_1$ BC this becomes, with $m={5 \over 2}f-{3 \over 2}L-{1 \over 4}$,
\begin{equation} 
\{ \sigma_1 V_m,  \varepsilon V_m, \sigma_2 V_m \},
\end{equation}
in accordance with the parafermion fusion rules $\sigma_1 \psi_1=\varepsilon$ and $\sigma_1 \psi_2=\sigma_2$.
For open/$\sigma_2$ BC, with $m={5 \over 2}f-{3 \over 2}L+{1 \over 4}$, 
\begin{equation}
\{ \sigma_2 V_m, \sigma_1 V_m, \varepsilon V_m \},
\end{equation}
in agreement with $\sigma_2 \psi_1=\sigma_1$ and $\sigma_2 \psi_2=\varepsilon$.
The supersymmetric groundstates are $|\sigma_{1,2} V_{\pm {1 \over 4}}\rangle$ and $|V_{\pm {3 \over 4}}\rangle$.  

Putting $\sigma_i$ type BC on both ends, the open-chain CFT sectors follow the fusion rules $\sigma_1 \sigma_1 = \psi_1+\sigma_2$,  $\sigma_1 \sigma_2 = 1+\varepsilon$ and  $\sigma_2 \sigma_2 = \psi_2+\sigma_1$. As for $k=2$, these fusion rules lead to exponential groundstate degeneracies in the extreme staggering limit. 
A characteristic case would be $L=15$ sites, $\sigma_2$-type BC on both ends, and staggering positioned as $\star\star\lambda \ldots$. Here the lowest CFT states are two supersymmetry doublets (at $f=8,9$) at CFT energies $E_{\rm CFT}=1/5$, $E_{\rm CFT}=4/5$, while the $\lambda=0$ $M_3$ model has four $E=0$ vacua. The states connect according to
\begin{eqnarray}
&| \frac{1}{2} \rangle_{\sigma_2,\sigma_2} \leftrightarrow |\psi_1 V_{-5/4}\rangle, & | 0 \rangle_{\sigma_2,\sigma_2} \leftrightarrow  |\psi_2 V_{+5/4}\rangle ,
\nonu
& |-\frac{1}{2} \rangle_{\sigma_2,\sigma_2} \leftrightarrow |\sigma_2 V_{-5/4}\rangle, & |1\rangle_{\sigma_2,\sigma_2} \leftrightarrow |\sigma_1 V_{+5/4}\rangle .
\label{k=3topostates}
\end{eqnarray}

\begin{figure}
   \subfloat[$\sigma_2$ defect]{
     \includegraphics{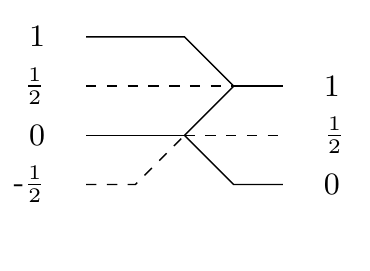}}
        \quad
      \subfloat[$\sigma_1$ defect]{
     \includegraphics{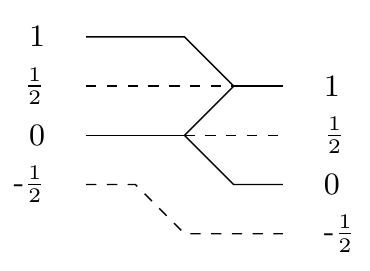}
      }
      \caption{\small{Fusion rules of bulk defects in the $M_3$ model at extreme staggering.}}
      \label{fig:fusion}
      \end{figure}

We define bulk defects as
\begin{eqnarray}
& \sigma_1 : \ \ldots  \underbracket{\lambda\star\star}_{{\rm no}\ 111} \ldots,\quad
& \sigma_2 : \ \ldots \lambda \underbracket{\star\ \star}_{{\rm no}\ 11}\  \ldots,
\nonu
& \sigma_1^\prime : \ \ldots \underbracket{\star\star\lambda}_{{\rm no}\ 111} \ldots,\quad
& \sigma_2^\prime : \ \ldots\  \underbracket{\star\ \star}_{{\rm no}\ 11}\lambda \ldots.
\end{eqnarray}
Fig.~\ref{fig:fusion} depicts the corresponding zero-energy fusion rules. They determine the size and structure of the quantum register opened up by an alternating sequence of defects $\sigma_i$, $\sigma_j^\prime$. In all cases the states at maximal particle number are made up entirely of vacua $|0\rangle$ and $|1\rangle$. Restricting the fusion rules to $|0\rangle$, $|1\rangle$ (drawn lines) gives Fibonacci number degeneracies. The ${\cal N}=2$ supersymmetry acts within the register, with $Q^-$ mapping the $|0\rangle$, $|1\rangle$ into linear combinations of $|\frac{1}{2}\rangle,|-\frac{1}{2}\rangle$.  In the example of $L=5l$ sites, $\sigma_2$/$\sigma_2$ BC, staggering type $\star\star\lambda\ldots$ and $n$ $\sigma_2$/$\sigma_2^\prime$-type defects, the degeneracy at $f=3l$ is ${\rm Fibo}_{n+3}$, with ${\rm Fibo}_j=1,1,2,3, \ldots$. Eq.~(\ref{k=3topostates}) is the case $n=0$. For $n=1$ one finds $3,4,1$ states at $f=3l,\ldots,3l-2$, $n=2$ gives $5,9,5,1$ states at $f=3l,\dots,3l-3$, etc.

For the general M$_k$ model, defects eliminating $k+1-j$ consecutive `1's will correspond to the $\mathbb{Z}_k$ parafermion spin fields $\sigma_j$, $j=1,\ldots,k-1$ \cite{FS15}. 

\vskip 3mm

{\bf Conclusions.\ -}\ We have demonstrated that (boundary and bulk) defects in the M$_k$ models off-criticality lead to quantum registers that are protected by supersymmetry. It will be interesting to explore ways to manipulate these registers, so as to act on the quantum information that can be stored in the supersymmetry protected ground states. One idea is that of a 1D braid protocol (as in \cite{alicea:11}); one expects that this will result in the braid matrices as they are known for the corresponding 2D (RR$_k$) topological phases. It will also be interesting to see 
if a dual formulation of the M$_k$ models, with 1D topological order rather than order of CDW type, can be obtained. 
One expects that operators that preserve supersymmetry in the M$_k$ models correspond to operators that are local in the dual variables of the topological phase, and are thus unable to split the exponential degeneracies.

Many extensions of the ideas presented here are feasible. In the M$_2$ model, alternative bulk defects, based on a `no 0' rather than a `no 11' condition, lead to Fibonacci number degeneracies. The M$_1$ model on a square ladder is in many ways similar to the M$_2$ model \cite{huijse:10} and it is natural to introduce $\sigma$-type defects in the 2D M$_1$ model on the octagon-square and square lattices \cite{fendley:05_2, huijse:10_2}. In all these cases, we expect non-trivial fusion relations to emerge.

\acknowledgements
We thank Erez Berg, Tarun Grover, Yingfei Gu, Christian Hagendorf, Liza Huij\-se, Steve Simon, and, in particular, Paul Fendley and Ville Lahtinen, for discussions, and the INFN for hospitality in Firenze, where part of this work was done. T.F. is supported by the Netherlands Organisation for Scientific Research (NWO).

\end{document}